# Sensor-aided block matching algorithm for translational motion estimation through a depth map

Karim El Khoury[1], Pascal Pellegrin[2], Antonin Descampe[2], Sébastien Lugan[1] and Benoit Macq[1]

[1]ICTEAM, UCLouvain, Louvain-la-Neuve, 1348, Belgium
[2]intoPIX S.A., Mont-Saint-Guibert, 1435, Belgium

**Abstract**

A large number of cameras embedded on smart-phones, drones or inside cars have a direct access to external motion sensing from gyroscopes and accelerometers. On these power-limited devices, video compression must be of low-complexity. For this reason, we propose a "Sensor-Aided Block Matching Algorithm" which exploits the presence of a motion sensor synchronized with a camera to reduce the complexity of the motion estimation process in an inter-frame video codec. Our solution extends the work previously done on rotational motion estimation to an original estimation of the translational motion through a depth map. The proposed algorithm provides a complexity reduction factor of approximately 2.5 compared to optimized block-matching motion compensated inter-frame video codecs while maintaining high image quality and providing as by-product a depth map of the scene.

## 1. Introduction

Inter-frame video compression, which exploits temporal redundancies, allows reaching very high-compression ratios via motion estimation [1]. Motion estimation relies on the estimation of motion vectors through block matching algorithms. Block matching consists of dividing the frame into non-overlapping blocks and localizing these blocks in the reference frame using a comparison criterion such as the Mean Squared Error, Mean Absolute Difference or Peak Signal-to-Noise Ratio [2]. A motion vector describes the displacement of each block in the frame. Improved block matching algorithms such as the Three-Step-Search [3], Four-Step-Search [4], Simple and Efficient Search [5], Diamond Search [6], and Adaptive Root Pattern Search [7] aim to reduce the computational complexity compared to the exhaustive Full-Search approach.

The main assumption of this paper is that further complexity reduction can be obtained by the use of gyroscopes and accelerometers, which are often available on platforms integrating mobile cameras. These new kinds of motion estimation could easily fit on any electronic device.

The proposed Sensor-Aided Block Matching Algorithm described in Fig. 1, aims to improve all of the previously listed algorithms by reducing the motion vector search through the use of an external motion sensor leading to a significant reduction of the computational complexity. It has been experimentally integrated in a low-power hybrid algorithm, for which the intra-frame and inter-frame residues compression are based on the low-latency and low-complexity JPEG XS standard [8] [9].

The rest of the paper is organized as follows. Section 2 presents the relation to prior work in rotational motion estimation. Section 3 provides the proposed enhanced approach integrating translational motion estimation via a depth map. Section 4 presents the depth map generation and motion vector prediction procedures. Section 5 provides the depth map visualization as well as performance results of the algorithm in terms of complexity gain and image quality. Section 6 concludes the paper.

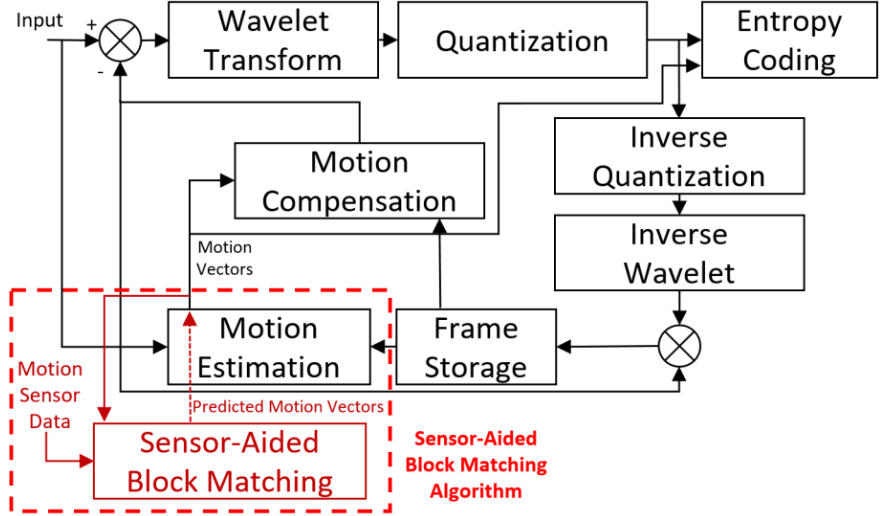

**Fig. 1.** Proposed hybrid video encoder block diagram.

## 2. Relation to prior work: rotational motion estimation

Estimating motion vectors from sensors attached to the camera has been explored by Hong et al. [10] and Chen et al. [11] for the compensation of rotational camera movements. These two contributions use two accelerometers placed on the camera to emulate a gyroscope.

Referring to Fig. 2, the object ($P$) is projected on the image plane as $p_1$. When the camera lens is rotated by an angle ($\theta$), the new projection of $P$ is denoted as $p_2$ on the new image plane. To get the image movement (motion vector) of $P$ generated by the rotational movement, we simply get the difference between the two projections of $P$ on both image planes. Therefore, the motion vector ($\Delta p$) is obtained using the following expressions, where $f$ is the focal length and $\alpha$ is the angle of view:

$$p_1 = f \cdot tan(\alpha) \quad (1)$$

$$p_2 = f \cdot tan(\alpha + \theta) \quad (2)$$

$$\Delta p = p_2 - p_1 = f \cdot \tan(\alpha + \theta) - f \cdot \tan(\alpha) \quad (3)$$

If we consider $\theta$ to be small enough as it is the angle of rotation between two consecutive frames we get:

$$\theta\ small \rightarrow \tan(\alpha + \theta) - \tan(\alpha) \approx \theta \cdot \sec^2(\alpha) \quad (4)$$

$$\Delta p \approx f \cdot \sec^2(\alpha) \cdot \theta \approx f \cdot \theta \quad (5)$$

where $sec^2(\alpha) \approx 1$ given that $\alpha$ is small enough in most camera lenses (except for wide-angle and fisheye lenses).

Therefore, the motion vector only depends on two parameters: the focal length (*f*), which is a constant specific to the camera, and the rotation angle ($\theta$). A synchronized motion sensor can be used to get this rotation angle and compute the motion vectors without requiring block matching. A gyroscope would give us the horizontal and vertical angular velocities. The rotational angles are computed by integrating those velocities. We can then calculate the horizontal and vertical predicted motion vectors, $\Delta p_h$ and $\Delta p_v$ respectively, and combine the two to give us the motion vector ($\Delta p$). Using this method, we can only compute the global movement (rotation of the camera) and not the local movement (objects moving in the scene) since the motion sensor can only account for movements of the camera.

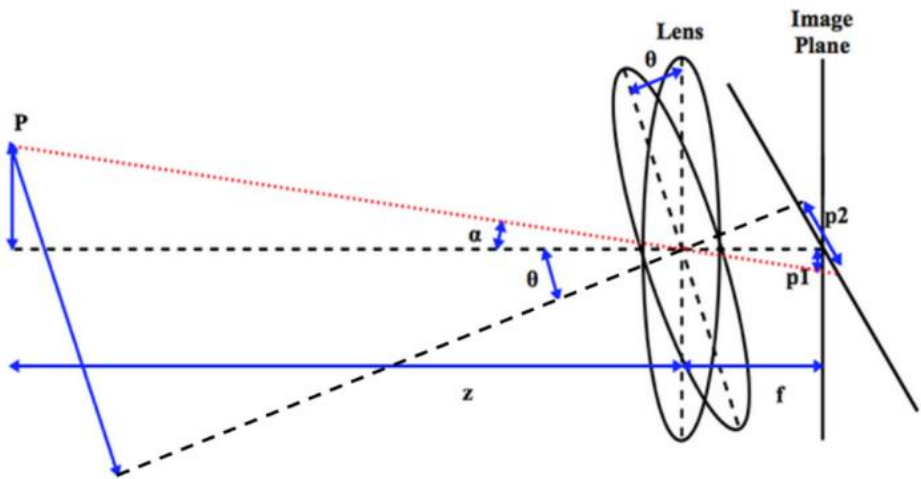

**Fig. 2.** Projective geometry for rotational camera movement.

## 3. Proposed approach: translational motion estimation through a depth map

The movement of a camera could be expressed as a combination of a rotation and a translation. Our proposed Sensor-Aided Block Matching Algorithm extends the work previously done [10] [11] on rotational camera movements to the more challenging case of translational movements. Translational motion estimation has already been explored in a specific way for deblurring [12].

Motion vectors produced by a translational movement can be retrieved from the accelerometer data through the computation of a depth map of the scene. The translation of the camera affects the various parts of the image depending on their specific depth level.

Referring to Fig. 3, the translation of the camera horizontally or vertically by $\pm d$ means that the new object is at position $P \pm d$ with respect to the new camera lens center. The motion vector ($\Delta p$) that effects the object ($P$) in the image is essentially the difference between the new projection ($p_2$) of $P+d$ on the image plane and the original projection ($p_1$) of $P$ on the image plane. $p_2$ and $p_1$ can be obtained using the following expressions, where $f$ is the focal length, $d$ is the translational displacement of the camera, and $z$ is the depth:

$$p_1 = \frac{f \cdot P}{z} \tag{6}$$

$$p_2 = \frac{f \cdot (P+d)}{z} \tag{7}$$

$$\Delta p = p_2 - p_1 = \frac{f \cdot d}{z} \tag{8}$$

The focal length ($f$) is a known parameter relative to the camera that is easily found. The translation displacement ($d$) in both the horizontal ($d_h$) and vertical ($d_v$) directions can be found by integrating the accelerometer data twice in both directions. The only variable that is unknown in the equation of the approximated motion vectors is the depth ($z$). The generation of the depth map of an inanimate scene is detailed in Section 4. Once we have the depth ($z$), we calculate the horizontal and vertical motion vectors ($\Delta p_h$) and ($\Delta p_v$) and combine them to get the motion vector ($\Delta p$).

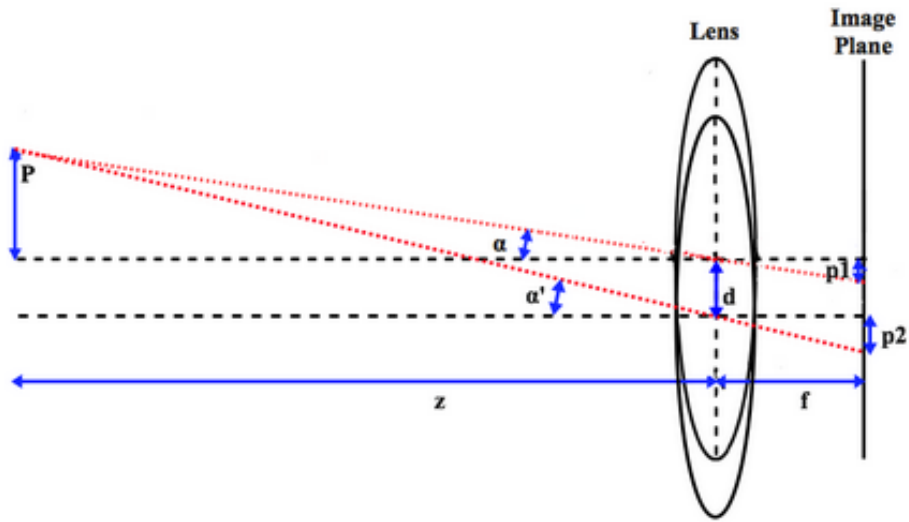

**Fig. 3.** Projective geometry for translational camera movement.

## 4. Depth Map Generation and Motion Vector Prediction procedures

The detailed block diagram of the Sensor-Aided Block Matching Algorithm is presented in Fig. 4. The training phase illustrated in blue represents the depth map generation. The prediction phase illustrated in orange represents the motion vector prediction.

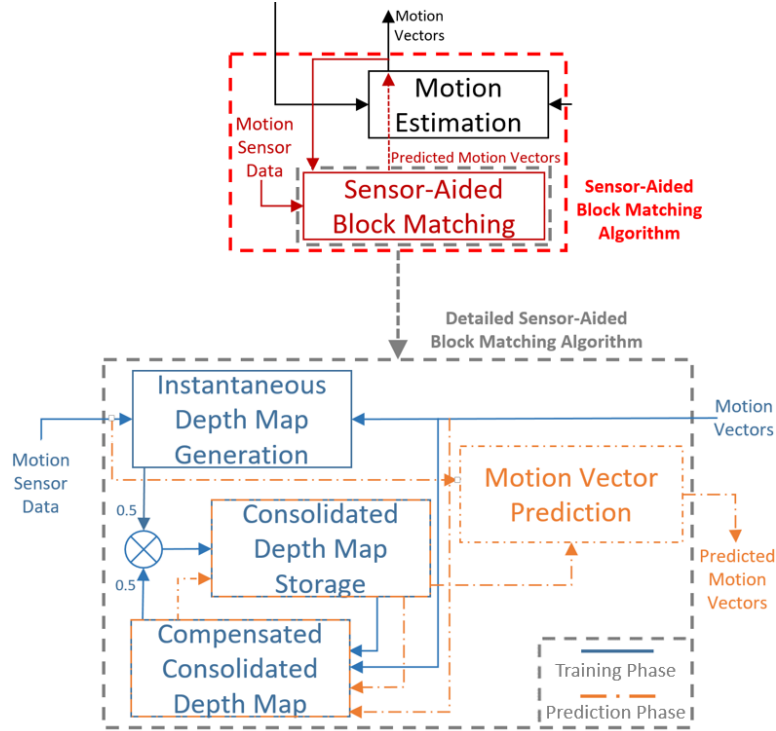

**Fig. 4.** Sensor-aided block matching detailed block diagram.

### 4.1. Depth Map Generation (Training Phase)

Essentially, the depth in the image is inversely proportional to the motion vector norms (refer to eq. 8). Therefore, we propose to define a training phase during which we first generate a depth map using the motion sensor data and the motion vectors (generated by classical block matching). We then consolidate this depth map by averaging it to increase its reliability. The training phase is described hereafter:

1. Generation of motion vectors at [t] via classical block matching between [t] and [t+1].
2. Calculation of the camera translational movement at [t] by double integration on accelerometer data between [t] and [t+1].
3. Calculation of the instantaneous depth map (for each frame) at [t] using eq. 8.
4. For the 1$^{st}$ frame, the consolidated depth map at [t] is the first instantaneous depth map at [t].
5. Calculation of the instantaneous depth map at [t+1] using eq. 8.
6. Generation of the consolidated depth map at [t+1] by shifting the consolidated depth map at [t] by the motion vectors at [t] and averaging it with the instantaneous depth map at [t+1].
7. This procedure repeats until the end of training phase.

### 4.2. Motion Vector Prediction (Prediction Phase)

The training phase is followed by a prediction phase which uses the consolidated depth map and the motion sensor data to estimate future motion vectors (refer to eq. 8). This process reduces the calculations compared to a classical block matching approach. The complexity gain is detailed in Section 5. The prediction phase is described hereafter:

1. Calculation of the camera translational movement at [t] by double integration on accelerometer data between [t] and [t+1].
2. Calculation of the predicted motion vectors between [t] and [t+1] using eq. 8 via the consolidated depth map at [t] and the camera translational movement at [t].
3. Smoothing of the predicted motion vectors via classical block matching (using a smaller sized search window).
4. Generation of the consolidated depth map at [t+1] by shifting the consolidated depth map at [t] by the motion vectors at [t] and averaging it with the instantaneous depth map at [t+1].
5. This procedure repeats until the end of prediction phase.

## 5. Results

The setup for the proposed algorithm has been tested using a 4K camera at 60 frames per second. It is synchronized with a motion sensor that extracts both accelerometer and gyroscope data. The setup is implemented on a Xilinx ZCU102 development board.

### 5.1. Depth Map Visualization

The experimental sequences used are characterized by translational movements in scenes with diverse depth levels to ensure the robustness of the algorithm. Fig. 5 shows the different stages of the depth map generation and consolidation procedures for the given scene in Fig. 5 (a). The instantaneous depth maps shown in Fig. 5 (b) and Fig. 5 (c) are generated using block matching with macroblocks of size 64x64 and 4x4 respectively. Combining both sizes is crucial for the precision of the produced motion vectors as well as the generation of smoother instantaneous depth maps as shown in Fig 5 (d). The final consolidated depth map at the end of the training phase is shown in Fig 5 (e).

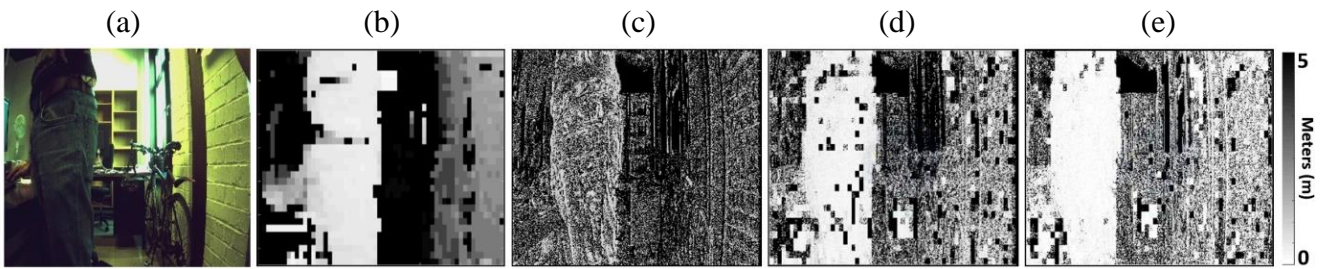

**Fig. 5.** (a) Given scene (b) Instantaneous depth map with 64x64 macroblocks (c) Instantaneous depth map with 4x4 macroblocks (d) Instantaneous depth map with 64x64 followed by 4x4 macroblocks (e) Consolidated depth map with 64x64 followed by 4x4 macroblocks.

## 5.2. Complexity gain

As previously stated in Section 5.1, the depth map generation during the training phase requires both large and small macroblocks (64x64 and 4x4). The Three-Step-Search algorithm [3] was chosen after evaluating several classical block matching algorithms. After experimentation, the search windows have been defined with search parameters of 63 and 15 for the large and small macroblocks respectively. The training phase consists of the depth map generation as well as the motion estimation. This motion estimation is performed using Three-Step-Search for large macroblocks followed by a small correction using the Three-Step-Search with small macroblocks. The prediction phase consists of the motion vector prediction followed by a small correction using the Three-Step-Search with small macroblocks. For block matching using the Three-Step-Search with a search parameter (*p*), the number of search points per macroblock (*sp/mb*) is calculated as follows:

$$sp/mb = 8 \cdot log_2(p+1) + 1 \qquad (9)$$

If the Mean Squared Error is used as our comparison criteria, we can calculate the multiplications needed for the Three-Step-Search per frame at 4K resolution by multiplying:

{a} the number of search points per macroblock (determined by eq. 9), {b} the macroblock size (multiplications needed by the Mean Squared Error) and {c} the total number of macroblocks in a 4K resolution frame. As for the depth map generation and motion vector prediction, the total number of multiplications needed per frame is calculated by the total number of macroblocks per frame (as each macroblock requires only one multiplication using eq. 8). This leads to the results detailed in Tab. 1, highlighting the complexity reduction aspect. A reduction factor of approximately 2.5 in terms of multiplications needed per frame is observed between the training phase and the prediction phase.

|  |  | Average complexity cost | Total average complexity cost |
|---|---|---|---|
| **Training Phase** | *Large Blocks (64x64)* | 385 351 680 | 645 365 760 |
|  | *Small Blocks (4x4)* | 259 522 560 |  |
|  | *Depth Map Generation* | 491 520 |  |
| **Prediction Phase** | *Motion Vector Prediction* | 491 520 | 260 014 080 |
|  | *Small Blocks (4x4)* | 259 522 560 |  |

**Tab. 1.** Average complexity cost comparison (multiplications needed per frame) of training phase versus prediction phase.

### 5.3. Image quality

The Sensor-Aided Block Matching Algorithm reasonably preserves the average PSNR performance at several entropy restriction levels between the training phase and the prediction phase as shown in Fig. 6. This highlights that the depth map generation and the motion vector prediction procedures are successful. As a result, the Sensor-Aided Block Matching Algorithm provides a complexity reduction compared to a classical inter-frame algorithm while preserving high image quality.

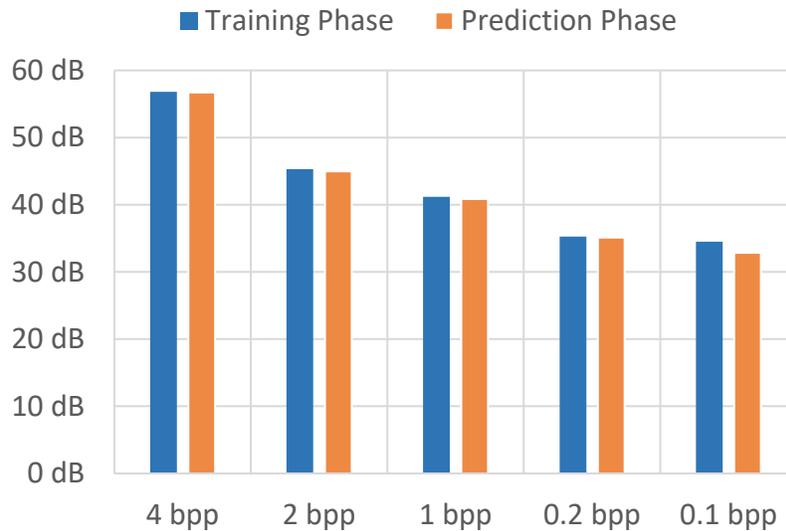

**Fig. 6.** Average PSNR at several entropy restriction levels (bit per pixels) of training phase versus prediction phase.

### 6. Conclusion

This paper proposes an innovative alternative to classical inter-frame codecs by utilizing the motion sensor data synchronized to the camera to decrease the complexity of the motion estimation process. We have experimented our proposal on 4K video sequences with a hybrid codec based on JPEG-XS integrating the Sensor-Aided Block Matching Algorithm. The results show significant complexity reduction results while maintaining high image quality. Given that motion sensors are embedded into readily available electronic devices, our algorithm is suitable to many potential applications besides video compression such as motion deblurring [12] [13]. This algorithm also provides a depth map by-product that benefits several use cases such as 3D reconstruction of scenes and obstacle detection [14] [15].

### 7. Acknowledgment

This research is supported by the TrustEye project (WALinnov program) funded by the Walloon region of Belgium and intoPIX S.A.